\documentclass[11pt,moriond,epsfig]{article}

\usepackage{moriond,epsfig,setspace}

\def\Journal#1#2#3#4{{#1} {\bf #2}, #3 (#4)}

\def\NIMA{{\em Nucl. Instrum. Methods} A}

\def\PLB{{\em Phys. Lett.}  B}
\def\PRL{\em Phys. Rev. Lett.}

\def\APP{\em Astropart. Phys.}
\def\SP{\em Sol. Phys.}
\def\ApJ{\em ApJ}

\def\be{\begin{equation}}
\def\ee{\end{equation}}
\def\bea{\begin{eqnarray}}
\def\eea{\end{eqnarray}}

\begin{document}
\vspace*{4cm}
\title{ANALYSIS OF EXPERIMENTS EXHIBITING TIME-VARYING NUCLEAR DECAY RATES: SYSTEMATIC EFFECTS OR NEW PHYSICS?}

\author{ J.H. JENKINS }
\address{School of Nuclear Engineering, Purdue University, \\400 Central Dr., West Lafayette, Indiana 47907 USA \\jere@purdue.edu}
\author{E. FISCHBACH}
\address{Department of Physics, Purdue University, West Lafayette, Indiana, 47907 USA}
\author{P.A. STURROCK}
\address{Center for Space Science and Astrophysics, Stanford University, Stanford, California 94305  USA}
\author{D.W. MUNDY}
\address{Department of Radiation Oncology Physics, Mayo Clinic, Rochester, Minnesota 55905 USA}
\maketitle\abstracts{
Since the 1930s, and with very few exceptions, it has been assumed that the process of radioactive decay is a random process, unaffected by the environment in which the decaying nucleus resides. There have been instances within the past few decades, however, where changes in the chemical environment or physical environment brought about small changes in the decay rates. But even in light of these instances, decaying nuclei that were undisturbed  or un-``pressured'' were thought to behave in the expected random way, subject to the normal decay probabilities which are specific to each nuclide. Moreover, any ``non-random'' behavior was assumed automatically to be the fault of the detection systems, the environment surrounding the detectors, or changes in the background radiation to which the detector was exposed. Recently, however, evidence has emerged from a variety of sources, including measurements taken by independent groups at Brookhaven National Laboratory, Physikalisch-Technische Bundesanstalt, and Purdue University, that indicate there may in fact be an influence that is altering nuclear decay rates, albeit at levels on the order of $10^{-3}$. In this paper, we will discuss some of these results, and examine the evidence pointing to the conclusion that the intrinsic decay process is being affected by a solar influence.
}

\section{Introduction}\label{intro}
It has long been a universal belief that radioactive decay is a random process, one that is almost completely 
insensitive to external influences. There have been a few special cases in recent times where minor changes in
decay rates have been measured due to artificially produced changes in the physical environment of the 
decaying nuclides\cite{eme72,hop74,hah76,dos77,nor01,oht04,lim06}, but on the whole the 
assumption has been that radioactive
decays follow the standard exponential decay law which is based on these decays 
being a random process. In recent years, however, a few independent groups have
identified some interesting behaviors in measured nuclear 
decay rates that did not arise from a change in the physical or chemical environment 
of the decaying nuclei.\cite{ell90,fal01,par05,bau07,par10a,par10b,shn98a,shn98b} In these results, there appears to be some
structure in what should be randomly distributed data points. More recently, however, 
Recent work by our group\cite{jen09a,jen09b,fis09,stu10a,jav10,stu10b,stu11a,stu11b} has gone further and
detailed the existence of periodicities and other
non-random behaviors in measured nuclear decay data from Purdue University\cite{jen09a}, Brookhaven
National Laboratory (BNL)\cite{alb86}, and the Physikalisch-Technische Bundesanstalt (PTB)\cite{sie98}.
The suggestion of this recent work is that there is a solar influence on the these measured decay rates,
via some particle or field of solar origin such as solar neutrinos.

Such a proposal is, without question, going to generate criticism from the physics 
community, based on the belief that the observed effects were the result of
changes in the environment of the detector systems (i.e., temperature, background, etc.) or 
systematic effects.\cite{coo09,nor09,sem09,sil09} However, a thorough analysis by our group of the
Purdue, BNL and PTB detector systems has effectively refuted
essentially all of this criticism.\cite{jen10} In this report we will further strengthen this view by
providing additional perspective and results that support the conjecture that whatever is influencing
the measured decay rates is external to the terrestrial environment, and could in fact have a solar origin.
 
\section{Review of Experimental Evidence}\label{expres}

To begin this discussion, it is helpful to collect together the information related to the observed
decay rate changes from multiple independent experiments. Table \ref{tab:exp}
lists several experiments which utilize different isotopes as well as different different detector
technologies, all of which show anomalous behaviors, either in the form of periodicities, or a localized departure
from the expected decay trend over a short duration.

\begin{table}[ht]
\caption{Experiments exhibiting time-dependent decay rates.\label{tab:exp}}
\vspace{0.4cm}
\scriptsize
%\onehalfspacing
\begin{center}
\begin{tabular}{|c|c|c|c|l|}
\hline
%\onehalfspacing
% & Detector & Experiment & Effect &\\
Isotope \&, & Detector & Radiation Type & Experiment & Effect \\
Decay Type &  Type & Measured & Duration & Observed \\
%&
%$\Gamma(\pi^- \pi^0)\; s^{-1}$ &
%$\Gamma(\pi^- \pi^0 \gamma)\; s^{-1}$ &
\hline
  &   &  &  &   \\
$^{3}$H, $\beta^{-}$ & Photodiodes &$\beta^{-}$ & 1.5 years & \textit{freq}(1/yr)\cite{fal01}  \\
$^{3}$H, $\beta^{-}$ & Sol. St. (Si) &$\beta^{-}$ &  4 years & \textit{freq}($\sim$2/yr)\cite{lob99} \\
$^{36}$Cl, $\beta^{-}$ & Proportional &$\beta^{-}$ &  8 years & \textit{freq}(1/yr, 11.7/yr, 2.1/yr)\cite{jen09b,stu10a,stu11a} \\
$^{54}$Mn, $\kappa$ &  Scintillation  & $\gamma$ &  2.5 months & Short term decay rate decrease\cite{jen09a} \\
$^{54}$Mn, $\kappa$ &  Scintillation  & $\gamma$&  2.5 years & \textit{freq}(1/yr) \\
$^{56}$Mn, $\beta^{-}$ &  Scintillation &$\gamma$ &  9 years & \textit{freq}(1/yr)\cite{ell90} \\
$^{60}$Co, $\beta^{-}$ & Geiger-M\"{u}ller  &$\beta^{-}$,$\gamma$ &  4.5 years & \textit{freq}(1/yr)\cite{par10a,par10b} \\
$^{60}$Co, $\beta^{-}$ & Scintillation &$\gamma$ &  4 months & \textit{freq}(1/d, ~12.1/yr)\cite{bau07} \\
$^{90}$Sr/$^{90}$Y, $\beta^{-}$ & Geiger-M\"{u}ller  &$\beta^{-}$ &  10 years & \textit{freq}(1/yr, 11.7/yr)\cite{par10a,par10b} \\
%  &  &  &  &   \\
$^{137}$Cs, $\beta^{-}$ & Scintillation &$\gamma$ &  4 months & \textit{freq}(1/d, ~12.1/yr)\cite{bau07}\\
$^{152}$Eu, $\kappa$ & Sol. St. (Ge) &$\gamma$ &  $>$16 years & \textit{freq}(1/yr)\cite{sie98} \\
$^{226}$Ra, $\alpha,\beta^{-}$ & Ion Chamber &$\gamma$ &  $>$16 years & \textit{freq}(1/yr, 11.7/yr, 2.1/yr)\cite{jen09b,stu10a,stu11a} \\
  &   &  &  &   \\

%\mco{2}{|c|}{Process for Decay} & & \\
%\cline{1-2}
%$K^-$ &
%$1.711 \times 10^7$ &
%\begin{minipage}{1in}
%$2.22 \times 10^4$ \\ (DE $ 1.46 \times 10^3)$
%\end{minipage} &
%\begin{minipage}{1.5in}
%No (IB)-E1 interference seen but data shows excess events relative to IB over
%$E^{\ast}_{\gamma} = 80$ to $100MeV$
%\end{minipage} \\
\hline
\end{tabular}
\end{center}
\end{table}

What should be evident from the information presented in Table \ref{tab:exp} is that the ``problem'' of 
apparent non-random behavior in nuclear decay measurements is apparent in a number of different experiments. What will probably also become evident as time passes is that the effect is more widespread than even this list
indicates. A simple search of the literature reveals multiple instances of articles discussing the discrepancies
in nuclear decay measurements, particularly half-life determinations.\cite{beg01,chi07,woo90,woo96} It is interesting,
given recent advances in detector technology, and the precision with which we can make measurements 
in the present day, that there would be discrepancies as large as are observed to be present in nuclear decay data. However, if some of
these measurements are of $\beta^-$ decays that are affected by an influence external to the Earth, and this 
influence has a variable output, then the picture becomes a little clearer. It is imperative, though, to rule out
the possible terrestrial influences such as the detector systems themselves, or changes in the local 
environment (temperature, barometric pressure, relative humidity, or background radiation) that could play a role in
producing these effects in the measured decay rates. Therefore, new experiments should record local conditions carefully if they are not able to be controlled completely. 

Returning to Table \ref{tab:exp}, we can draw some conclusions about the possible influence of 
environmental and systematic effects from the list presented there. To begin, all of the isotopes 
presented in Table \ref{tab:exp} are $\beta$-decays, or $\beta$-decay related, even the $^{226}$Ra
measured on the ionization chamber at the PTB.\cite{sie98} Clearly, while $^{226}$Ra is 
not a $\beta$-decay itself, there are several $\beta$-decaying daughters in its decay chain, nearly 
all of which are in equilibrium with the $^{226}$Ra parent.\footnote{Good descriptions of the $^{226}$Ra 
decay chain and the equilibrium activities of a $^{226}$Ra source are presented by Christmas\cite{chr83} and Chiste {\em et al.}\cite{chis07}} Since the ionization chamber system utilized in the PTB 
experiment was not designed to differentiate between the specific photons emitted by the $^{226}$Ra or
or any of its daughters, it is impossible to determine whether the decay rate changes were occurring in 
the $\alpha$- or $\beta$-decays of the chain. No effects have been seen 
in $\alpha$-decays to this point,\cite{par10a,coo09,nor09} which is not surprising. Since 
the mechanisms of $\alpha$- and $\beta$-decays are so different, the fact that the effect has not been observed in
$\alpha$-decays should not exclude the possibility of the effect existing in $\beta$-decays.\cite{coo09}

Upon further examination of the experiments described in Table \ref{tab:exp}, we see that there is representation of all three major classes of detector types, solid state (2), scintillation (5), and gas detectors (4). There is also a mix of the types of radiation detected, about equally split between charged particles ($\beta^-$) and photons ($\gamma$). There is one experiment (the one presented by Falkenberg\cite{fal01}) in Table \ref{tab:exp} that is unique in that the detection method did not fit into any of the standard classes. The experiment utilized photodiodes to measure the radioluminescence of tritium tubes.

When examining the possible environmental influences on the radiation transport (from source to detector, which is in general over a very short distance on the order of a few millimeters in most cases), the primary consideration is the air density of the source-detector gap, which will be a function of temperature ($T$), barometric pressure ($P$), and relative humidity ($RH$). A thorough discussion of this is presented in Jenkins, Mundy and Fischbach,\cite{jen10} who note that cool, dry air is much more dense than warm, moist air. Interestingly, the effect seen in \textit{all} of the experiments listed in Table \ref{tab:exp} exhibit higher counts in the winter, when the air is ostensibly denser. If air density (as a function of $T, P, \rm{and}~RH$) is higher in the winter due to the air being cooler and drier, then the count rates of the charged particles should be lower in the winter due to the greater energy loss as the $\beta$-particles interact with more gas atoms in the denser air, not higher. The transport of photons across the small source-detector gaps will not be affected by air density at a level worth considering. Furthermore, a detailed analysis utilizing MCNPX performed by our group\cite{jen10} supports the above qualitative arguments, and thus refutes claims to the contrary by Semkow \textit{et al.}\cite{sem09} that the observed effects were strictly due to environmental influences on the detector systems.

The variety of detector systems helps to offset other possible environmental or systematic influences as well, since there are no known systematic effects that would affect each of the different systems in the same way. For instance, with the Geiger-M\"{u}ller detectors, a single ionization can cause an avalanche and ionize all of the gas that can be ionized within the entire tube, thus there is no pre-amplifier or amplifier required. This eliminates the opportunity for shifting in the electronics that would affect peak shape, or other similar properties of the system. One may reasonably draw the conclusion that there are no systematic effects which would be likely to have caused these periodicities. However, we can pursue that in yet another way.

Looking at the ``Observed Effect'' column in Table \ref{tab:exp}, we note there exists more than just an annual frequency in seven of the twelve experiments. We note here that the ``Observed Effects'' column list the frequencies discussed in the respective articles describing the experimental results (in one case, the $^{54}$Mn data that show an annual oscillation, the full frequency analysis has not yet been performed, these are new data presented for the first time here, see Section \ref{newres}). While there may be other frequencies present in these experimental data sets, those analyses are not available. What is important to remember, however, is that these frequencies are exhibited in data that should not have any frequency structure at all. While it may be easier to discard an annual frequency by attributing it to the change of the seasons, it is impossible to say the same about an approximate monthly frequency (which appears in five of the experiments) or a roughly semi-annual frequency (which appears in three, and two of those three also contained the monthly frequencies). It is also not likely that one could offer a systematic explanation for the existence of those periodicities. Therefore, it is reasonable to look outside the local laboratory conditions, making the solar influence certainly plausible.

\section{New Results}\label{newres}

\begin{figure}[ht]
\centering
\includegraphics[scale=0.75]{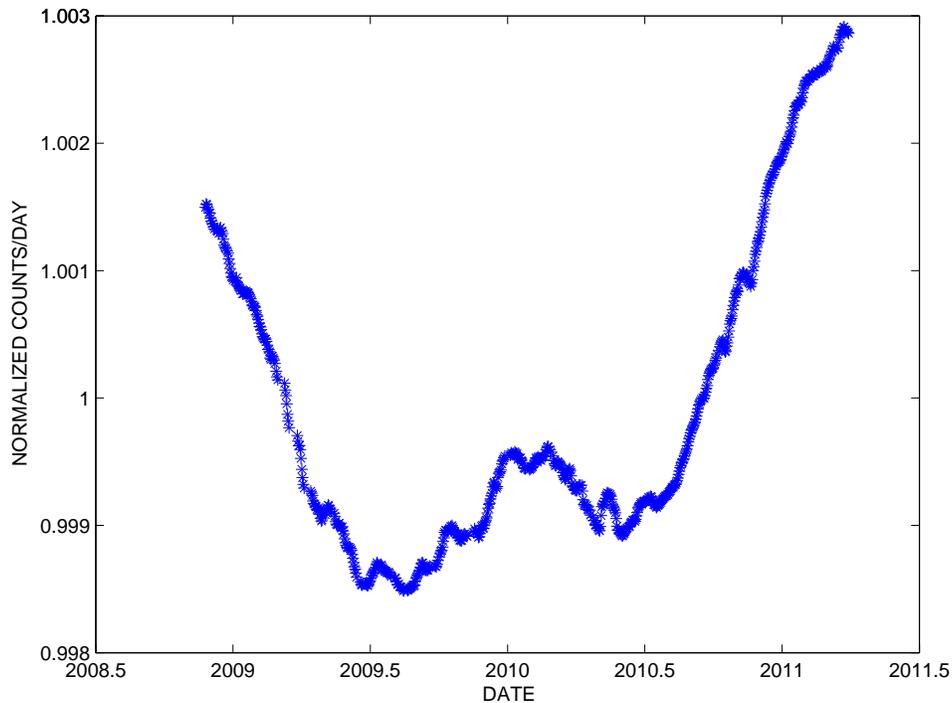} 
\caption{$^{54}$Mn decays measured at Purdue University. The 834.8 keV photon was measured with a 2-inch NaI detector, and were taken continuously for 3600 seconds live time, then aggregated into counts/day. These integrated counts were then undecayed (detrended), and normalized to the average of the series.
\label{fig:fig1}}
\end{figure}

%\begin{figure}[ht]
%\caption{Caption}
%\includegraphics[scale=1]{J-Mor_Fig1.eps}
%\end{figure}
In November 2008, our group began measuring $^{54}$Mn again, taking continuous, 3600 s live-time counts. The results of the measurement series are still preliminary, but we present an overview here. Each of the counts in a 24-hour period were aggregated into one data point which represents counts/day, then a 21-day sliding average centered on each point was calculated to smooth the data set in order to show long-period oscillations more clearly. The results are shown in Figure \ref{fig:fig1}. The presence of a frequency with a period of one year is obvious, as is the indication of some shorter period frequencies. A detailed analysis will be presented in a forthcoming paper. This annual frequency is also listed in the data presented in Table \ref{tab:exp}.

\begin{figure}[ht]
\centering
\includegraphics[scale=0.85]{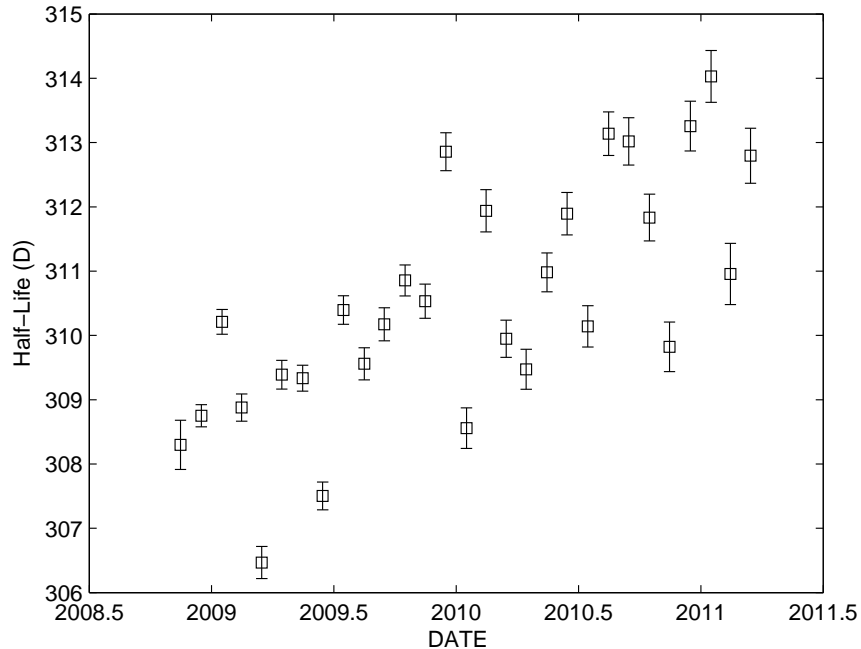} 
\caption{Variation in the measured $^{54}$Mn half-life, looking at one-month segments of the decay measurement series. The half-life value for each month was calculated by performing a weighted least squares fit to an average of 664 data points, with the $\sqrt{N}$ fractional uncertainty of each point varying from $\sim0.03\%$ at the beginning of the experiment to $\sim0.07\%$ near the end. The average $\chi^2/d.o.f$ for each fit was $\sim1.3$.
\label{fig:fig2}}
\end{figure}

The presently accepted half-life of $^{54}$Mn is 312.12(6) days,\cite{jun06} and from our data we have determined the half-life to be 310.881(2) days. Our data set contains 19,191 separate 3600 s live-time counts over 877 days (2.81 half-lives) totalling 1.01$\times10^{11}$ measured decay events. What is curious is that the $\chi^2$/d.o.f of the weighted least-squares fit is 7.99, which is fairly large. However, after examining the plot in \ref{fig:fig1}, the fact that the data are not distributed randomly around the value 1.00, and stray from that normalized value of 1, raises an interesting question: How does the half-life vary in shorter segments of the entire set, which is a question similar to the one examined by Siegert, Schrader and Sh\"{o}tzig\cite{sie98} for $^{152}$Eu. We have calculated the half-life for each month of data by performing a weighted least squares fit to an average of 664 data points per month, with the $\sqrt{N}$ fractional uncertainty of each point varying from $\sim0.03\%$ at the beginning of the experiment to $\sim0.07\%$ near the end. These monthly half-lives are shown in \ref{fig:fig2}. The average $\chi^2/d.o.f$ for each month's fit was $\sim1.3$, which is a great improvement over the fit to the whole line. It is easy to see that there is a fairly significant variability in the measured count rates. We measured the environmental conditions in the laboratory ($T,~P~and~RH$) and these were found not to vary significantly, and also did not correlate with the variability in the measured count rates. A more rigorous analysis is under way, the results of which will be available soon.

In light of all of this evidence, it seems clear that all of the possible, known systematic effects or environmental effects are too small to have caused the oscillatory or other ``non-random'' characteristics in the data from the experiments listed in Table \ref{tab:exp}. Without question, more work needs to be done in determining what the cause is, but based on all of the evidence presented by our group, it appears that the most likely external influence at this time is the Sun. It is, therefore, our hope that many new experiments will be undertaken by groups around the world to continue this work. Even if the cause turns out to not be solar-related, identifying and understanding this effect will have a broad impact across the world of science and technology related to nuclear decays.

\section*{Acknowledgments}
The work of PAS was supported in part by the NSF through Grant AST-06072572, and that of EF was supported in part by U.S. DOE contract No. DE-AC02-76ER071428.

\end{document}